\documentstyle[prd,aps,epsfig,floats]{revtex}

\newcommand{\dfrac}{\displaystyle \frac}

\begin{document}
\draft
\twocolumn[\hsize\textwidth\columnwidth\hsize\csname
@twocolumnfalse\endcsname

\title{Prediction of Three-body $B^0\rightarrow
\rho^-p\bar{n},\ \pi^-p\bar{n}$ Decay Rates}

\author{Chun-Khiang Chua, Wei-Shu Hou and Shang-Yuu Tsai\\}
\address{Department of Physics, National Taiwan University,
Taipei, Taiwan 10764, R.O.C.}
\date{\today}
\maketitle

\begin{abstract}

We predict the rates of the charmless three-body
$B^0\to \rho^-p\bar{n}$ and $\pi^-p\bar{n}$ modes due to
weak vector current contributions to be
$\sim 4\times 10^{-6}$ and $2\times 10^{-6}$, respectively. The
basis is a factorization approach of current produced nucleon
pairs, together with an isospin transformation that relates
nucleon weak vector form factors to electromagnetic form factors.
Adding the axial vector current contribution, we find $B^0\to
\rho^-p\bar{n}$ and $B^+\to \rho^0p\bar{n}$ to be at $10^{-5}$
order. The three-body modes appear to dominate over
the two-body modes such as $B\to p\bar{p}$, $p\bar \Lambda$.

\end{abstract}

]
%
%
%
%


\section{Introduction}\label{Intro}

A large number of charmless mesonic decays of the $B$ mesons
have emerged since 1997, and are of great current interest.
For example, $\pi^+\pi^-/K^+\pi^-\sim 1/4-1/5$ suggests~\cite{HHY}
that $\phi_3$ (or $\gamma$) $\equiv{\rm arg}\,V_{ub}^*$
could be $90^{\circ}$ or more.
A natural question to ask \cite{Hou:2001bz} is:
what about charmless baryonic decays?
The CLEO Collaboration has done some search in the past,
but turning up null results for modes like $B^0\to p\bar{p}$
that are below the $10^{-5}$ level \cite{CLEObaryo}.
The Belle Collaboration has recently improved the limits \cite{ppbar}
on $B^0\to p\bar p$ by an order of magnitude,
pushing down to the $10^{-6}$ level.
It is of interest to ask, therefore, if all charmless
baryonic modes  are below $10^{-5}$.
In this paper, from a suitably sound theoretical
basis involving nucleon form factor data,
we show that $B^0\to \rho^-p\bar{n}$ could be a leading charmless
baryonic decay with rate at the $10^{-5}$ level.

The CLEO Collaboration recently reported the observation of
the $B^0\rightarrow D^{*-}p\bar{n}$ mode at the $10^{-3}$ level
\cite{Anderson:2001tz}, which is only a factor of 4--5 lower than
$B^0\rightarrow D^{*-}\rho^+$ and $D^{*-}\pi^+$~\cite{Groom:2000in}.
Scaling by $\vert V_{ub}/V_{cb}\vert^2$ one could already
infer that $B^0\rightarrow \rho^{-}p\bar{n} \sim 10^{-5}$,
but a better understanding is desirable.
A factorization approach for $B^0\rightarrow D^{*-}p\bar{n}$ with
current produced nucleon pairs has been proposed recently
\cite{PreviousWork}.
The three-body decay is seen as generated by two weak currents:
one converting $B^0$ to $D^{*-}$, the other creating the nucleon pair.
The nucleon weak vector form factors are related
by isospin rotation to nucleon electromagnetic form factors.
By using $G^{p,n}_M$ measured from $e^+e^-\rightarrow \bar{N}N$
and $p\bar{p}\rightarrow e^+e^-$ processes
\cite{timelikeData,Armstrong93,Antonelli:1998fv,Ambrogiani:1999bh},
we are able to account for up to $60\%$ of the observed rate,
the remainder seemingly coming from axial vector current contribution.
Emboldened by this success, we apply the approach to
the charmless modes $B^0\to \rho^-p\bar{n},\ \pi^-p\bar{n}$
where one replaces the $D^{*-}$ by $\rho$ or $\pi$.

Besides rates, we are able to predict the $p\bar n$ pair mass spectrum.
Since the vector current contributes dominantly to the total rate
for the $B^0\to D^{*-}p\bar{n}$ case~\cite{PreviousWork},
we expect the vector current to dominate
the $B^0\to\rho^-p\bar{n}$ and $\pi^-p\bar{n}$ rates as well.
Incorporating estimates of the axial current contributions,
the total rates are slightly higher than from the vector current alone.

\section{Formalism}\label{formalism}

\begin{figure}[b!]
\begin{center}
\epsfig{figure=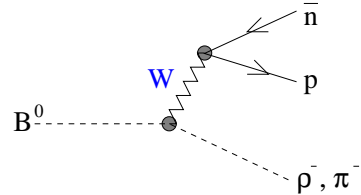,width=1.83in}
\end{center}
\caption{Feynman diagram illustrating
Eq.~(\ref{factorization}).}
\label{Feyn_fact}
\end{figure}

Our starting point is to
factorize the current production of $p\bar{n}$ pairs, i.e.
\begin{eqnarray}
&&\langle \rho^-(\pi^-) p\bar{n}|{\cal H}_{eff}|B^0\rangle
=\frac{G_F}{\sqrt{2}}\,V_{ud}V_{ub}^*\,a_1\, \nonumber\\
&&\quad\quad\quad\quad \times
\langle \rho^-(\pi^-)|V^{\mu}-A^{\mu}|B^0\rangle 
\langle p\bar{n}|V_{\mu}-A_{\mu}|\,0\rangle\,.
\label{factorization}
\end{eqnarray}
The $V-A$ current induces
the $\bar{b}\to \bar{u}$ transition in the first term,
while in the second it creates the nucleon pair.
which is illustrated in Fig.~\ref{Feyn_fact}.
It is an extension of the usual factorization
of current-current matrix elements in the case of $B$ decays to 
two mesons \cite{Wirbel:1985ji,Ali:1998eb}.

Once factorized, the matrix element
$\langle p\bar{n}|V_{\mu}-A_{\mu}|\,0\rangle$
describes nucleon pair creation by a charged weak current.
The matrix element for the vector ($V_{\mu}^+$) portion
can be expressed as
\begin{equation}
\langle p\bar{n}|V_{\mu}^+|0\rangle= \bar{u}(p_p)
\left\{F_1^W(t)\gamma_{\mu}+
   i\,\frac{F_2^W(t)}{2m_N}\sigma_{\mu\nu}q^{\nu}\right\}v(p_{\bar{n}}),
\label{weak_current}
\end{equation}
where $m_N$ is the nucleon mass,
$q\equiv (p_p+p_{\bar n})$ the momentum transfer,
$t\equiv q^2 = m_{p\bar{n}}^2$ the $p$-$\bar n$ invariant mass squared,
and $F_{1,2}^W$ are nucleon weak form factors,
with $F^W_1(t)$ normalized at $t=0$~\cite{Georgi:1984kw},
\begin{equation}
F^W_1(0)=1.\label{norm1}
\end{equation}

The photon field $A_\mu$ contains $W^3_\mu$, which,
together with $W^{1,2}_\mu$, form a weak isotriplet.
The coupled currents also form an isotriplet, and can be
interrelated via an isospin transformation. 
For the nucleon pair, the strong isospin symmetry of the nucleon
state coincides with the weak isospin symmetry of the weak and em
currents. The weak vector form factors are therefore related to
electromagnetic (em) isovector form factors.

The matrix element $\langle N(p')\bar{N}(p)|{\cal J}^{em}_{\mu}|0\rangle$
for the em current can be expressed as
\begin{eqnarray}
&&\langle N(p')\bar{N}(p)|{\cal J}^{em}_{\mu}|0\rangle
\nonumber\\
&& =\bar{u}(p') \biggl\lbrace F_1 (t)\gamma_\mu + i\,
\frac{F_2(t)}{2m_N} \sigma_{\mu \nu}  q^\nu \biggr\rbrace v(p),
\label{em-current}
\end{eqnarray}
where $F_{1,2}(t)$ are respectively the Dirac and Pauli form factors,
normalized at $t=0$ as
\begin{equation}
F_1^p(0) = 1, \quad F_1^n(0) = 0, \quad 
F_2^p(0) = \kappa_p, \quad F_2^n(0) = \kappa_n,
\label{norm2}
\end{equation}
with $\kappa_{p\ (n)}$ the proton (neutron) 
anomalous magnetic moment in nuclear magneton units. 
These form factors are related to the Sachs form factors via
\begin{eqnarray}
G^{p,n}_E(t)&=&F_1^{p,n}(t)+\frac{t}{4m_N^2}F_2^{p,n}(t)\,,
\nonumber\\
G^{p,n}_M(t)&=&F_1^{p,n}(t)+F_2^{p,n}(t)\,.
\label{EMff}
\end{eqnarray}
The isospin decomposition of the em current is given by
\begin{equation} F_i^{s,\,v} = \dfrac{1}{2}\, (F_i^p
\pm F_i^n) \, , \quad 
\quad i = 1,2 \, , \label{isospin_decomp}
\end{equation}
where $s,\ v$ stand for the isoscalar and isovector components,
respectively. The isovector component of the em current and the
vector portion of the charged weak currents form an
isotriplet, as manifested by~\cite{Georgi:1984kw}
\begin{equation}
2\,F_i^{v}(t)=F_i^W(t),\quad\quad\quad\,\, i=1,2, \label{WeakEMff}
\end{equation}
where the factor $2$ is from the definition of
$F_{1,2}^{(s,v)}(t)$ in Eq. (\ref{isospin_decomp}).
For example, from Eqs.~(\ref{norm1}), (\ref{norm2}) 
and (\ref{isospin_decomp}) one easily checks that 
$2\,F_1^{v}(0)=F_1^W(0)$.

We can now write the three-body
$B^0\to \rho^- p \bar n$ decay amplitude in the following form
\begin{eqnarray}
&&
i{\cal{M}}_V=  
\left(-i\,\frac{G_F}{\sqrt{2}}
V_{ud}V^*_{ub}\,a_1\right)\,\epsilon_{\rho}^{*\nu}\,
\Biggl[-\epsilon_{\mu\nu\alpha\beta}
\,
p^{\alpha}_B p^{\beta}_{\rho}
 \frac{2V(q^2)}{m_B+m_{\rho}}
 \nonumber\\
&& \quad \
-ig_{\mu\nu}(m_B+m_{\rho})A_1(q^2)
+i\left(p_B+p_\rho\right)_\mu\,q_\nu\frac{A_2(q^2)}{m_B+m_\rho}\Biggr]
\nonumber\\
&& \quad
\times \bar{u}(p_p)
\Biggl[2\left(F^{v}_1+F^{v}_2\right) \gamma^{\mu}
 +\frac{F^{v}_2}{m_N} \left(p_{\bar{n}}-p_p\right)^{\mu}
\Biggr]v\left(p_{\bar{n}}\right),
 \label{amplitude}
\end{eqnarray}
and for $B^0\to \pi^- p \bar n$,
\begin{eqnarray}
&&i{\cal{M}}_P=  
\left(-i\,\frac{G_F}{\sqrt{2}}
V_{ud}V^*_{ub}\,a_1\right)\left(p_B+p_{\pi}\right)_{\mu} F_1(q^2)
\nonumber \\
&&
 \quad
\times \bar{u}(p_p)
\Biggl[2\left(F^{v}_1+F^{v}_2\right) \gamma^{\mu}
 +\frac{F^{v}_2}{m_N} \left(p_{\bar{n}}-p_p\right)^{\mu}
\Biggr]v\left(p_{\bar{n}}\right),
\label{amplitude2}
\end{eqnarray}
where $\epsilon_{\rho}$ is the $\rho$ meson polarization,
and $V(q^2)$, $A_1(q^2)$, $A_2(q^2)$ and $F_1(q^2)$
(not to be confused with baryon form factors)
are the transition from factors arising from
the $\bar{b}\to\bar{u}$ transition in
the first matrix element of Eq.~(\ref{factorization}).

As we need to integrate over $q^2$ for these three-body decay
modes, we need to pay more attention to the $q^2$ dependence of these 
transition form factors \cite{Cheng:1999kd}.
However, since our focus is on utilizing experimental data
on baryon form factors, for $B\to\rho,\pi$ form factors,
we shall simply take what is readily available in the litetature. 
Among the several recent models for the meson form factors 
(see e.g. Ref. \cite{Braun:1999,hep-ph/0001113}), 
we shall use~\cite{hep-ph/0001113}
\begin{equation}
f(q^2)=\frac{f(0)}{(1-q^2/M_V^2)(1-\sigma_1 q^2/M_V^2)},
\end{equation}
for $F_1(q^2)$ and $V(q^2)$, and
\begin{equation}
f(q^2)=\frac{f(0)}{1-\sigma_1 q^2/M_V^2+\sigma_2 q^4/M_V^4},
\end{equation}
for $A_{1,2}(q^2)$. $M_V$ is the appropriate pole mass which is
taken to be 5.32~GeV. 
Note that the $q^2$ dependence is quite different from the monopole
form used in Ref. \cite{Wirbel:1985ji}.
In Table~\ref{poles} we give the values of
the relevant form factors at zero momentum transfer as well as the
parameters $\sigma_{1,2}$~\cite{hep-ph/0001113}.

%
\begin{table}[t!]
\begin{center}
\caption{Form factors at $q^2 = 0$ and the parameters
$\sigma_{1,2}$.}
\begin{tabular}{cllll}
 & $V^{B\rho}$  & $A^{B\rho}_1$ & $A^{B\rho}_2$ & $F^{B\pi}_1$ \\
\hline
$f(0)$     & 0.31 & 0.26 & 0.24 & 0.29 \\
$\sigma_1$ & 0.59 & 0.73 & 1.40 & 0.48 \\
$\sigma_2$ & \,\,--- & 0.10 & 0.50 & \,\,--- \\
\end{tabular}
\label{poles}
\end{center}
\end{table}

It is important to note that the baryon form factors must satisfy
perturbative QCD~(PQCD) quark counting rules~\cite{Brodsky:1975vy},
which give the leading power large-$|t|$ fall-off of
the $F^v_1(t)$ form factor.
Since helicity-flip gives an extra $1/t$ factor for $F^v_2(t)$,
one finds in the large $|t|$ limit
\begin{eqnarray}
F^v_i (t) \to (|t|)^{-(i+1)} \biggl[
\ln\biggl(\dfrac{|t|}{Q_0^2}\biggr) \biggr]^{-\gamma} &,&\quad
 \,\quad i = 1,2\,, \label{asym}
\end{eqnarray}
where $Q_0 \simeq \Lambda_{\rm QCD}=0.3$ GeV, 
$\gamma=2+4/(3\beta)$,
and $\beta$ is the QCD $\beta$--function to one loop.
We note that $\gamma$ depends weakly on the number of flavors;
for three flavors $\gamma=2.148$.
The asymptotic form given in Eq.~(\ref{asym}) has been confirmed
by many measurements of the nucleon form factors
$G^{p,n}_M=F^{p,n}_1+F^{p,n}_2$ over a wide range of momentum
transfers in the space-like region \cite{eeScattering}.
It has also been confirmed in
the time-like region with the recent nucleon time-like
data~\cite{Armstrong93,Antonelli:1998fv,Ambrogiani:1999bh}.

The combination $2(F^v_1+F^v_2)$ in Eq.~(\ref{amplitude}) can be
replaced by $G^p_M-G^n_M$ which is composed of measurable quantities.
Similar replacement can also be made for $F^v_2$,
which is a combination of $G^p_M-G^p_E$ and $G^n_M-G^n_E$.
Most time-like data for the magnetic form factors, however, are
extracted by assuming either $|G^N_E|=|G^N_M|$ or $|G^N_E|=0$ in the
explored region of momentum transfer. Since $G^N_M - G^N_E = (1
-{t}/{4m_N^2}) \, F^N_2$ clearly vanishes at threshold,
by assuming $|G^N_E|=|G^N_M|$ in extracting $G^N_M$ from data, the
information on $F^N_2$ is lost.
In our calculation we concentrate on
the part of Eq.~(\ref{amplitude}) which contains $F^v_1+F^v_2$,
the contribution from $F^v_2$ can be
determined only when $G^N_M$ and $G^N_E$ can be
separated from data with better angular resolution.

\begin{figure}[t!]
\begin{center}
\epsfig{figure=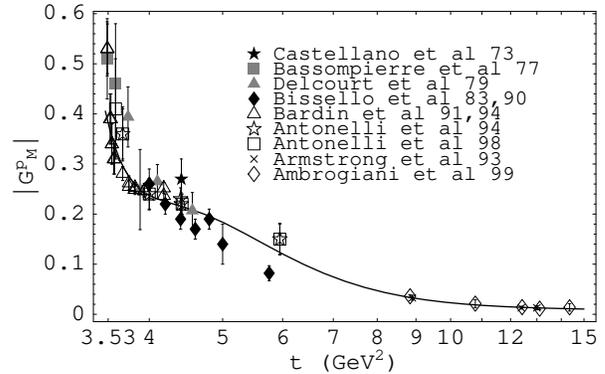,width=3.3in}\end{center}
\caption{Time-like proton magnetic form factor data,
fitted by Eq.~(\ref{fitP})
with the parameters given in Eq.~(\ref{bestP}).
}
\label{proton}
\end{figure}

\begin{figure}[t!]
\begin{center}
\epsfig{figure=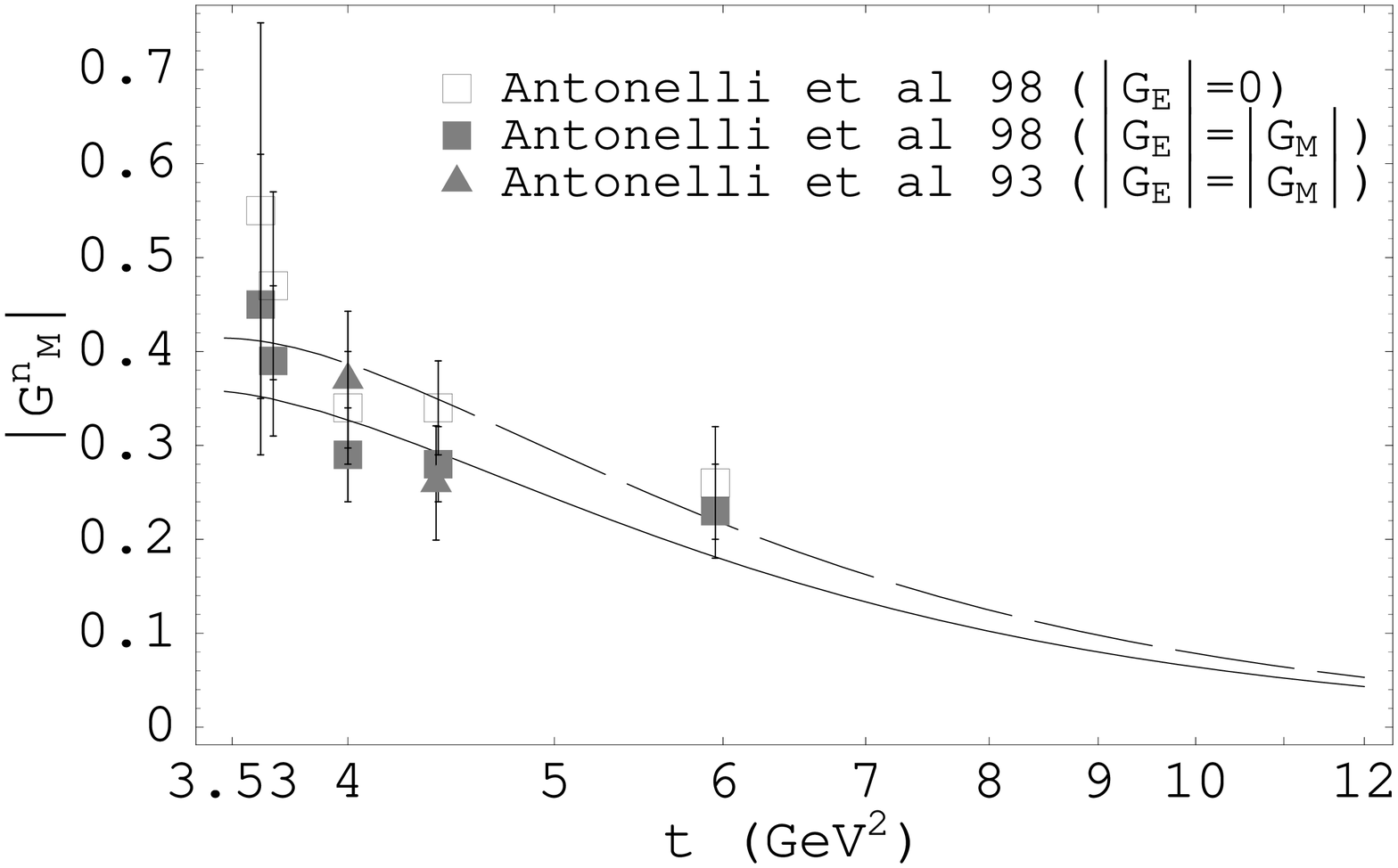,width=3.3in}
\end{center}
\caption{ Time-like neutron magnetic form factor, where the solid
(long-dash) line is the fit given by Eq.~(\ref{fitN}) with parameters
given in Eq.~(\ref{bestNe}) (Eq.~(\ref{bestN})),
for data extracted with $|G^n_E|=|G^n_M|$ ($|G^n_E|=0$) assumption.
}
\label{neutron}
\end{figure}

We take $|G^N_M|$ in the following form~\cite{PreviousWork} to make
a phenomenological {\it fit} of the experimental data
\cite{timelikeData,Armstrong93,Antonelli:1998fv,Ambrogiani:1999bh}:
\begin{equation}
\left|G^p_M(t)\right|= \sum^5_{i=1}\frac{x_i}{t^{i+1}}
\left[\ln\left(\frac{t}{Q_0^2} \right)\right]^{-\gamma},
 \label{fitP}
\end{equation}
\begin{equation}
\left|G^n_M(t)\right|=\sum^2_{i=1}\frac{y_i}{t^{i+1}}
\left[\ln\left(\frac{t}{Q_0^2}\right)\right]^{-\gamma},
 \label{fitN}
\end{equation}
the power of the leading term and logarithmic factor are as
suggested by PQCD, and the fewer number of fit parameters for
$G_M^n$ reflects the fact of scarcer neutron data. We find the
best fit values
\begin{eqnarray}
&x_1&=429.88~\mbox{GeV}^4\,,
\,\,\,\quad\quad\quad x_4=-448583.96~\mbox{GeV}^{10}\,,\nonumber\\
&x_2&=-10783.69~\mbox{GeV}^6\,,
\quad\quad x_5=635695.29~\mbox{GeV}^{12}\,, \nonumber\\
&x_3&=109738.41~{\rm GeV}^8,
 \label{bestP}
\end{eqnarray}
and
\begin{equation}
y_1=236.69~{\rm GeV}^4\,,\quad \quad \quad y_2=-579.51~{\rm
GeV}^6\,, \label{bestNe}
\end{equation}
where the $\chi^2$ per degree of freedom (d.o.f.) of the fits are
$1.39$ for $|G^p_M|$ and $0.41$ for $|G^n_M|$, respectively.
We show in Figs.~\ref{proton} and \ref{neutron} the best fit curves
given by Eqs.~(\ref{fitP}) and (\ref{fitN}) with the above
parameters.

It was pointed out in Ref.~\cite{Antonelli:1998fv} that the data
supports $|G^n_E|=0$ as well. We therefore perform a fit to
the neutron magnetic form factor data extracted
under the assumption of $|G^n_E|=0$, giving the best fit values
\begin{equation}
y_1=292.62~{\rm GeV}^4\,,\quad \quad \quad y_2=-735.73~{\rm
GeV}^6\,, \label{bestN}
\end{equation}
with $\chi^2/$d.o.f.$=0.39$, which is slightly lower than the previous
fit. This fit is also plotted in Fig. \ref{neutron}.
More data is needed to distinguish between the two cases.
%

We note that there is a sign difference between $G^p_M$ and
$G^n_M$ in the space-like region. Since 
analyticity implies continuity at infinity between space-like and
time-like \cite{Logunov} regions,
the time-like magnetic form factors are expected to
have similar behavior as the space-like ones:
real and positive for the proton, but negative for the neutron.

For large $t$, QCD predicts the magnetic form factors to be
real~\cite{Brodsky:1975vy}, with the neutron form factor weaker
than the proton case~\cite{Matveev73}. According to QCD sum
rules~\cite{Chernyak:1984bm}, asymptotically one expects
$G^n_M/G^p_M\sim Q_d/Q_u=-0.5.$ In our fits, with
sign difference between $G^p_M$ and $G^n_M$, we have
$G^n_M/G^p_M=-y_1/x_1=-0.55~(-0.68)$ for
$|G_E|=|G_M|~(|G^n_E|=0).$
Nucleon form factors have also been analyzed from negative to
positive $t$ with dispersion relations. The phase of the proton
magnetic form factor turns out to be $\sim 2\pi$, hence the
proton magnetic form factor is real and positive as expected
asymptotically, starting already from $t\ge
4$~GeV$^2$~\cite{Hammer:1996kx,Baldini:1999qn} onwards.

\section{Results and Discussion}

We still need to fix $V_{ud} V_{ub}^* a_1$.
We shall take $|V_{ud}|=0.9747$ and $|V_{ub}|=34.95\times
10^{-4}$ from Ref. \cite{Ciuchini:2000de}.
For the effective coefficient $a_1$,
we take the value $a_1=1.05$ from Ref.~\cite{Ali:1998eb}
for effective number of color $N_c=3$.
The situation is slighly different from previous study on
$B^0\to D^{*-} p \bar n$ case \cite{PreviousWork}, where $a_1$
is taken from $B^0\to D^{*-} \rho^+$ decay \cite{Cheng:1999kd}.
Here, when one tries to follow the procedure by looking at 
$B^0\to\rho^-\rho^+,\,\pi^-\rho^+$ modes, the tree-penguin interference
will complicate things \cite{HHY,Hou:2000tf}.
We therefore use the short distance $a_1$ for simplicity. 

For proton and neutron data extracted assuming $|G^{p,n}_E|=|G^{p,n}_M|$,
we find the branching ratio for
$B^0\to \rho^- p\bar{n}$ arising from the vector current is
\begin{equation}
\mbox{Br}_V(B^0\to \rho^- p\bar{n})=(3.58^{+0.70}_{-0.61}) \times
10^{-6}\left(\frac{a_1}{1.05}\right)^2,
 \label{BSWBr}
\end{equation}
%
where the subscript $V$ is a reminder that this is from the vector
portion of the weak current alone. The upper and lower bounds
correspond respectively to the maximum and minimum of the
branching fraction evaluated by scanning through
$\chi^2\leq\chi^2_{\rm min}+1$ in the fits.
In a similar fashion, for data extracted assuming
$|G_E^p|=|G_M^p|$ but $|G_E^n|=0$, we find
\begin{equation}
\mbox{Br}_V(B^0\to \rho^- p\bar{n})=(4.53^{+1.03}_{-0.88}) \times
10^{-6}\left(\frac{a_1}{1.05}\right)^2.
\label{BSWBr0}
\end{equation}

Following the same methods, we also give predictions on the
$B^0\to \pi^- p\bar{n}$ rate that arise from the vector current.
We find
\begin{equation}
\mbox{Br}_V(B^0\to \pi^- p\bar{n})=(1.82^{+0.17}_{-0.16}) \times
 10^{-6}\left(\frac{a_1}{1.05}\right)^2 ,
\end{equation}
for $|G^{p,n}_M|=|G^{p,n}_E|$, and
\begin{equation}
\mbox{Br}_V(B^0\to \pi^- p\bar{n})=(2.29^{+0.49}_{-0.42}) \times
10^{-6}\left(\frac{a_1}{1.05}\right)^2 ,
\end{equation}
for $|G^{p}_E|=|G^{p}_M|$ but $|G^n_E|=0$.

\begin{figure}[t!]
\begin{center}
\epsfig{figure=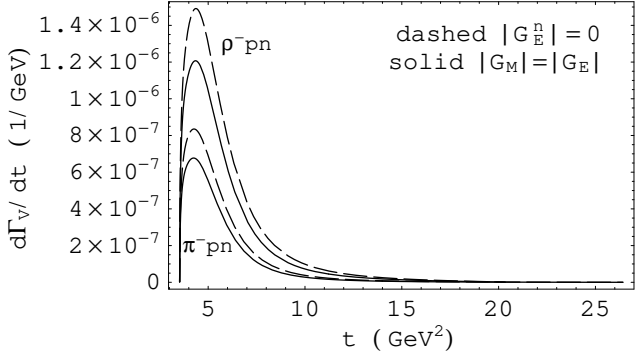,width=3.3in}
\end{center}
\caption{ The vector current induced differential decay rates
${d\Gamma_V(B^0\rightarrow \rho^-p\bar{n})}/{dq^2}$ (upper two curves)
and ${d\Gamma_V(B^0\rightarrow \pi^-p\bar{n})}/{dq^2}$ (lower two curves). 
Solid lines are from fitting nucleon form factor
data with $|G^{p,n}_M|=|G^{p,n}_E|$, dashed lines with
$|G^{p}_M|=|G^{p}_E|$ and $|G^n_E|=0$. }
\label{dGdqVrhopi}
\end{figure}

Fig.~\ref{dGdqVrhopi} shows the vector current induced
differential decay rates ${d\Gamma_V}/{dq^2}$
of both the $B^0\to\rho^-p\bar{n}$ and the $\pi^-p\bar{n}$ modes
with BSW form factors. The peaking of the differential rates at
$\sim 5$ GeV$^2$ and $\sim 4$ GeV$^2$ for the $\rho^-p\bar{n}$ and
$\pi^-p\bar{n}$ modes, respectively, is a threshold enhancement
effect for baryon production. As argued in \cite{Hou:2001bz}, a
fast recoil meson carries away energy and
would be more favorable for baryon production in the recoil system
because of reduced energy release compared to two-body decays.
This fast recoil meson accompanying
the low mass baryon pair can be tested experimentally.
We note that the narrowness of the $t$ distribution in
Fig.~\ref{dGdqVrhopi} 
implies the recoil meson spectrum of a quasi-two-body mode.

According to the $B^0\to D^{*-}p\bar{n}$ study~\cite{PreviousWork},
the vector current contributes $\sim 60\%$ of
the observed rate~\cite{Anderson:2001tz}.
We can estimate the total rates of
the $B^0\to \rho^-p\bar{n},\ \pi^-p\bar{n}$ modes
by assuming similar proportions of the vector current contributions.
The estimated rates are then
Br$(B^0\to\rho^-p\bar{n})\sim 7\times10^{-6}$ and
Br$(\pi^-p\bar{n})\sim 3\times10^{-6}$,
i.e. $B^0\to\rho^-p\bar{n}$ is of order $10^{-5}$.
Inspection of Fig. 1 suggests that
$B^+\to\rho^0 p\bar{n}$ is at the same order.

It is interesting to compare the three-body rates
Br$(B^0\to h p\bar{n})$ with the two-body
ones Br$(B^0\to h\rho^+)$ where $h$ stands for the recoil meson.
We find a similarity of the ratios Br(3-body)/Br(2-body)
between $h=D^{*-}$ and $h=\rho^-,\pi^-$.
By taking $\phi_3=54.8^{\circ}$~\cite{Ciuchini:2000de}, we obtain
Br$(\rho^-\rho^+(\pi^-\rho^+))
\sim 32~(22)\times 10^{-6}$~\cite{Hou:2000tf}.
We then have
Br$(\rho^-p\bar{n}(\pi^-p\bar{n}))/$Br$(\rho^-\rho^+
 (\pi^-\rho^+))
\sim
 0.22~(0.14)$ which is rather close to
Br$(D^{*-}p\bar{n})/$Br$(D^{*-}\rho^+)\sim$ 0.2.


So far, we have assumed only the
tree level $b\to d\bar uu$ transitions as the
underlying process, as illustrated in Eq. (1) and Fig. 1.
As we make comparison with charmless mesonic modes,
it is important to remember that,
unlike the $B^0\to D^{*-}p\bar{n}$ case,
we expect significant penguin contributions as well.
It is known that $b\to s\bar uu$ penguins dominate $K\pi$ modes,
and that penguin contributions make
significant impact on $\pi\pi$ modes.
Since the impact of penguins becomes
less pronounced for $VP$ and $VV$ modes,
and since our approach to $\rho^- p\bar n$ corresponds to
both $VP$ and $VV$ components,
we expect the impact of penguins on $B^0\to \rho^- p\bar n$
to be not as pronounced as in $B^0\to \pi^+\pi^-$.
On the other hand,
extending our discussion to $b\to s$ penguin
operators suggests three-body charmless baryonic modes such
as $\rho^-\Lambda\bar p$, $K^*\Lambda \bar\Sigma$ etc.
However, this would involve further assumptions.

With recent improvement of $B^0\to p\bar p$ limit to $10^{-6}$
level \cite{ppbar},
we now have the intriguing situation that
$B\to\rho p\bar{n} \gg B\to p\bar p$,
i.e. the three-body baryonic mode dominates over the two-body.
This would confirm the conjecture made in Ref. \cite{Hou:2001bz}.
We urge the Belle and BaBar groups to search
for the three-body modes experimentally.

In summary, from a relatively robust foundation,
in analogy with the $B^0\rightarrow D^{*-}p\bar{n}$ mode
observed by CLEO,
we have shown that $B^0\to\rho^-p\bar{n}$ is very likely
at the $10^{-5}$ level and should be well within
the capabilities at B Factories.
Together with the absence of two body modes such as
$B^0\to p\bar p$, the dynamics of $B$ decay to baryons
are somewhat different from mesonic final states.
Extending our study to include the standard set of effective operators
should be straightforward,
and one expects a host of three-body baryonic modes
at the $10^{-5}$ level.
We expect charmless baryonic modes
to emerge soon at the B Factories, likely from three-body onwards.

\acknowledgments

This work is supported in part by the National Science Council of
R.O.C. under Grants NSC-89-2112-M-002-063, NSC-89-2811-M-002-0086
and NSC-89-2112-M-002-062, the MOE CosPA project, and the BCP
topical program of NCTS.


\end{document}